\documentclass{aa}  

\usepackage{graphicx}
\usepackage{txfonts}
\usepackage{amsmath}
\begin{document}

   \title{Braking index of the frequently glitching PSR J0537$-$6910}

%  \subtitle{I. Overviewing the $\kappa$-mechanism}

   \author{Erbil G\"{u}gercino\u{g}lu \inst{1,2,8}, Onur Akbal \inst{3}, M.~Ali Alpar \inst{4}, Danai Antonopoulou \inst{5} \and Crist\'obal M. Espinoza \inst{6,7} }
   \institute{
          School of Arts and Sciences, Qingdao Binhai University, Huangdao District, 266555, Qingdao, People's Republic of China\\
             \email{egugercinoglu@gmail.com}
        \and
        National Astronomical Observatories, Chinese Academy of Sciences, 20A Datun Road, Chaoyang District, Beijing 100101, China\\ 
        \and 
        Bah\c{c}e\c{s}ehir College, \c{C}i\c{c}eklik\"{o}y, Bornova, 35040 Izmir, Turkey\\
              \email{akbalonur85@gmail.com}
         \and
        Faculty of Engineering and Natural Sciences, Sabanc{\i} University, Orhanl{\i}, Tuzla, 34956 Istanbul, Turkey\\
             \email{ali.alpar@sabanciuniv.edu}
         \and 
        Jodrell Bank Centre for Astrophysics, Department of Physics and Astronomy, The University of Manchester, UK\\
          \email{antonopoulou.danai@gmail.com}
          \and
        Departamento de F\'isica, Universidad de Santiago de Chile {\sc (USACH)},  Chile\\
          \email{cristobal.espinoza.r@usach.cl}
          \and
        Center for Interdisciplinary Research in Astrophysics and Space Sciences {\sc (CIRAS)}, {\sc USACH}, Chile\\
          \and
      Department of Astronomy and Space Sciences, Istanbul University, Beyaz{\i}t, 34119, Istanbul, Turkey\\
          }
   \date{...}
   \date{Erbil G\"{u}gercino\u{g}lu dedicates this paper to the beloved memory of his dear mother G\"{u}l\"{u}\c{s}an G\"{u}gercino\u{g}lu.}

% \abstract{}{}{}{}{}
% 5 {} token are mandatory
 
  \abstract
   {The pulsar J0537$-$6910 undergoes spin-up glitches more frequently than any other known pulsar, at a rate of roughly thrice per year. Its glitches are typically large and accompanied by abrupt changes of the spin-down $\dot{\nu}$ that partially recover with a nearly constant positive frequency second derivative $\ddot{\nu}$ for the entire post-glitch interval until the next glitch. The effective long-term value of $\ddot{\nu}$, however, is negative because $\dot{\nu}$ has decreased over the years of observations.}
  {We wish to determine if `permanent shifts' (non-relaxing parts of the glitch change $\Delta\dot{\nu}$ in the 
  spin-down rate, like those observed in the Crab pulsar) can explain the long-term enhancement of the spin-down rate which results in an effective negative braking index. We demonstrate, as a proof of concept, that the actual braking index associated with the pulsar's braking torque can be $n \sim 3$ if the internal superfluid torque and permanent shifts are considered.}
   {We use published RXTE and NICER data to calculate the average permanent shift per glitch needed to bring an underlying braking index $n$ 
   to the effective long-term value $n' \cong -1.2$ inferred from the data. We then use this average value as the actual 
   permanent shift in each glitch and extract the contributions of the internal and external torques to the observed $\ddot{\nu}$ for each 
   inter-glitch interval, under the assumption that the next glitch occurs when all glitch-induced offsets to internal torques are fully restored. }
   
  {We find that if the braking index of the magnetospheric torque is close to $n\sim3$, moderate permanent changes of the spin-down rate are required, of magnitude similar to the persistent shifts inferred for the Crab pulsar. The natural candidate mechanism to produce such permanent changes is that of crust-quakes. Crustal failure associated with PSR~J0537$-$6910 glitches can have interesting -- and potentially observable --  consequences, such as transient changes of the X-ray emission, activation of radio emission, or emission of gravitational waves.} 


\keywords{stars: neutron -- pulsars: general -- pulsars: individual: PSR J0537$-$6910}
\authorrunning{G\"{u}gercino\u{g}lu et al.}
\titlerunning{Braking index of PSR J0537$-$6910}
  
   \maketitle   

\section{Introduction}

The X-ray pulsar PSR J0537$-$6910 in the Large Magellanic Cloud (LMC) is exceptional in its rotational properties. Its spin evolution is interrupted by large glitches, i.e. abrupt changes in the rotation and spin-down rates, followed by post-glitch relaxation. 
[see reviews on pulsar glitches: \cite{haskell15,zhou22,antonopoulou22,antonelli22}]. This is the most frequently glitching known pulsar, with a quite regular rate of $\gtrsim 3.2$ glitches per year and glitch rotational frequency steps $\Delta\nu$ among the largest observed in young pulsars 
\citep{middleditch06,antonopoulou18,ferdman18,abbott21a,ho20,ho22}. 

The large glitch activity makes the measurement of the braking index, $n =\nu \ddot{\nu}/ \dot{\nu}^2$ (where $\nu$, $\dot\nu$ and $\ddot\nu$ are the pulse frequency and its first 
and second time derivatives, respectively), quite ambiguous. A pulsar slowing down purely under magnetic dipole braking would have $n=3$.
\citet{middleditch06}, \citet{antonopoulou18} and \citet{ho22} have determined an effective  \textit{negative} braking index from the overall increase of the spin-down rate (decrease in $\dot{\nu}$) in the long run, while \citet{antonopoulou18, ferdman18} have measured high inter-glitch braking indices, typically over $ n\sim 7$ and up to $n\sim 100 $ from data between successive glitches. These anomalously high values are dominated by the internal 
torques that couple the crust to the neutron star superfluid interior; separating their effects in the timing data allows one to find the underlying braking index due to the external 
braking torque. 

The paper is organized as follows: In Section \ref{timingbeh}, we overview the timing data and rotation of PSR J0537$-$6910. In Section \ref{intcoupling}, we discuss the post-glitch spin-down evolution of multi-component neutron stars with references to the vortex creep model.  
In Section \ref{sec:model}, we present the relationships between the observable quantities and the hidden undelying braking index, with the results for the model parameters detailed in Section \ref{sec:results}. Section \ref{glitchpre} discusses a method for estimating the time to the subsequent glitch, of potential  future use in scheduling an observation campaign covering the 
the next glitch in PSR J0537$-$6910. We discuss our results and their implications in Section \ref{sec:conc}.

\section{Timing Behavior of PSR J0537$-$6910}\label{timingbeh}

PSR J0537$-$6910, discovered with the Rossi X-ray Timing Explorer (RXTE) \citep{marshall98}, is located inside the 1 to 5 kyr old supernova remnant N157B \citep{chen06} in the Large 
Magellanic Cloud, 49.6 kpc away from Earth \citep{pietrzynski19}. It is the fastest rotating young pulsar, with a spin frequency $ \nu=62 $ Hz, and has a very high spin-down rate ($ \dot{\nu}=-1.992 \times 10^{-10}$ Hz s$^{-1}$), resulting in the largest known spin-down luminosity $ \dot{E}=4.9 \times 10^{38} $ erg s$^{-1}$. 
The characteristic age $\tau_{\rm c}=\nu/2\dot\nu$ is about $4.9 $ kyr and the surface dipole magnetic field $ B_{\rm s}$ is inferred to be $ \cong 9.25 \times 10^{11} $ G. The source is observed primarily in the X-ray band, with some weak radio emission detected recently \citep{crawford24}.

Glitches in PSR J0537$-$6910 were initially reported by \citep{marshall98,marshall04}.
Later, all RXTE observations available were analysed by \citet{antonopoulou18}, who reported a total of 45 glitches in the nearly 13 yr of data. 
Their findings were in agreement with \citet{ferdman18}, who also reports no evidence of pulse flux and profile changes associated with the glitches. 
After the decommissioning of RXTE in 2012, PSR J0537$-$6910 was observed with the Neutron Star Interior Composition Explorer (NICER) that discovered 15 glitches \citep{ho20,abbott21b,ho22}. Five further glitches in the Jodrell Bank Glitch Catalogue\footnote{https://www.jb.man.ac.uk/pulsar/glitches/gTable.html} are not included in our analysis as the timing data are not yet published for these recent glitches.

All PSR~J0537$-$6910 glitches have similar glitch steps in the rotation frequency ($\Delta\nu$) and the spin-down rate ($\Delta\dot\nu$), and similar relaxation properties. The glitch activity parameter, calculated as $A_{\rm g} \cong \sum_{j=1}^{N_{\rm g}} \left(\Delta\nu/\nu\right)_{\rm j}/T_{\rm obs} = 9 \times 10^{-7} {\rm yr}^{-1}$ 
where $T_{\rm obs}$ is the total time span containing $N_{\rm g}$ observed glitches, each with fractional spin increases $\left(\Delta\nu/\nu\right)_{\rm j}$. This is the 
second highest glitch activity parameter after PSR J1023--5746, which has $A_{\rm g}=14.5\times10^{-7}$ yr$^{-1}$ \citep{erbil21}. 

The magnitude of the spin-down rate $\left|\dot{\nu}\right|$ usually increases at glitches, so that $\Delta\dot{\nu}<0$. This step partially recovers with a nearly constant positive spin frequency 
second derivative $\ddot{\nu}$ until the time of the next glitch. 
This is typical inter-glitch timing behaviour in pulsars with large glitches.
For PSR J0537$-$5910, the inter-glitch spin-down rate evolves with $\ddot{\nu} \sim 10^{-20}$ Hz s$ ^{-2}$ on average, which implies an anomalously large average inter-glitch braking index
$ n_{\rm ig} \cong 20 $.
This value can be up to 5 times larger for some glitches. 
Exponentially relaxing components are not as prominent in PSR~J0537$-$6910 glitches as for other pulsars, e.g. Vela, however, after the first glitch in the RXTE data,
\citet{antonopoulou18} 
identified an exponentially decaying component with a time-scale
$\tau_{\rm d}=21\pm 4$ days followed by a linear spin-down regime with
an inter-glitch braking index $7.6\pm0.1$.
Exponential recoveries have recently been detected for 11 additional
glitches, most of which are followed by regimes with inter-glitch
braking indices between 6 and 9 (Zubieta et al. 2025, private
communication).  
\citet{ferdman18} applied Markov Chain Monte Carlo (MCMC)
techniques to simultaneously fit all RXTE inter-glitch $\dot{\nu}$
data by aligning them at the glitch epoch. 
They measured a decay time-scale $\tau_{\rm d}=27^{+7}_{-6}$ d and an
inter-glitch braking index $\sim7.4$.
Using likewise superposed inter-glitch data in terms of
$\ddot{\nu}(t)$ \citep{andersson18} or in terms of the short term
braking index \citep{ho20} yields time-scales $\tau_{\rm d}= 19 - 44$ d for the post-glitch exponential relaxations.

 The high glitch activity of this source makes its braking index measurements quite ambiguous. In the long-term, the evolution of the spin-down $\dot\nu(t)$ has the characteristic sawtooth shape seen in pulsars with regular large glitches, however the overall slope is negative. This translates into a long-term \textit{negative} apparent braking index $n' \equiv \ddot{\nu}_{\rm long-term}\nu_{\rm 0} / \dot{\nu}_{\rm 0}^2 \simeq  $ - 1.2 \citep{antonopoulou18,ho22}, where $\nu_{\rm 0}$  and $\dot{\nu}_{\rm 0}$ are the rotation frequency and spin-down rate, respectively, at some reference epoch and the long-term frequency second derivative $\ddot{\nu}_{\rm long-term}$ is measured for the entire data span, over many glitches. The long-term behaviour is likely the artefact of the high  
glitch activity, as discussed in \citet{antonopoulou18, ferdman18,ho22}.

\section{Spin-down of the multi-component neutron star with internal couplings}\label{intcoupling}

In steady-state, the crust and all interior components of the neutron star spin down at the same rate, dictated by the external (pulsar) braking torque and maintained 
by lags $\omega \equiv \Omega_{\rm s}- \Omega_{\rm c} >0$ between the rotation rates $\Omega_{\rm s}$ of the various interior (superfluid) components and the rotation rate $\Omega_{\rm c}$ of the outer crust (neutron star surface) and interior components co-rotating with the outer crust. In response to offsets from the steady-state lag introduced at each glitch (which decreases $\omega$), the observed spin-down rate of the crust experiences step increases at each glitch. Subsequently, the lag will evolve back to the steady-state value under the effect 
of the external (pulsar) torque, resulting in the relaxation of the observed crust spin-down rate \citep{alpar84,erbil20}. This is 
the typical inter-glitch timing behavior of pulsars with large glitches, of which the Vela pulsar is the prototype \citep{akbal17,erbil21}. Internal contributions to the 
glitch-induced change in the spin-down rate will eventually fully relax back to the pre-glitch steady-state. Such features are common to multi-component systems driven by external torques or 
forces. In addition, as a result of possible structural changes that affect the coupling of the superfluid to normal components, there can be permanent, non-relaxing changes associated with glitches, similarly to what is seen as `persistent shifts' in the spin-down rate of 
the Crab pulsar \citep{demianskiproszynski83, lyne92,lyne15}.

If the coupling between some internal superfluid component and the crust is linear in the lag $\omega$, then post-glitch relaxation will proceed
exponentially. The return to steady-state on the time-scale $\tau_{\rm d}$ of the exponential decay can be discerned in the post-glitch timing data of many 
pulsars (e.g., \citet{ yu13,lower21,erbil21,zubieta23,liu24,grover25, zubieta25}). On the other hand, if the coupling is non-linear, then the relaxation takes the form of a power-law, which has 
no scale; or of some more complicated response function which may involve `waiting times’ correlated with the offset introduced by the glitch, i.e. with the parameters of the glitch. 
Exponentials, power-laws and --occasionally-- more complicated features are common in the inter-glitch behaviour of pulsars, indicating the presence of both linearly and non-linearly coupled regions. 
The most common manifestation of non-linear coupling is a roughly constant second derivative $\ddot{\nu}$, called `anomalous’ because its value is too large to be 
associated with $n \sim 3$ dipole radiation torque, as the contribution of the internal torque is larger, and in many cases much 
larger, than that of the external pulsar torque \citep{alpar06}. Such signatures are common in glitching pulsars \citep{yu13,lower21,erbil21,zubieta24,liu24}, and studied most extensively 
in the Vela pulsar \citep{alpar84,akbal17,erbil21,grover25}. 

The scale-free power-laws arising from the internal torque can dominate the spin-down rate for an entire inter-glitch interval, and must be accounted for in order to find the contribution of the external 
torque and the actual braking index. A data set containing many glitches, as in the case of PSR J0537$-$6910, makes it possible to calculate  inter-glitch fits that extend all the way until the next glitch  \citep{antonopoulou18,ferdman18, ho20, ho22, abbott21b}. Ignoring for a moment the effect of the external torque and any permanent changes, the measured $\ddot{\nu}$ from such fits, together with the total increase of the spin-down rate observed 
at each glitch, give a recovery time-scale of $t_{\rm g}\cong |\Delta\dot{\nu}|/\ddot{\nu}$. This time-scale can be interpreted as the waiting time for return to steady-state inside the neutron star, when conditions will be suitable for another glitch to occur \citep{alpar84}.
 
The vortex creep model \citep{alpar84} explains the inter-glitch constant second derivative term quite naturally: 
In the presence of local pinning sites 
that hinder the free motion of vortices, the radially outward motion of vortex lines required to spin the superfluid down in response to the pulsar torque is achieved by thermal 
activation  of vortices over pinning barriers. The process is highly non-linear because of Boltzmann factors which depend exponentially on the lag $\omega = \Omega_{\rm s} - 
\Omega_{\rm c}$ between the superfluid rotation rate $\Omega_{\rm s}$ and the crust rotation rate $\Omega_{\rm c}$. In steady-state, vortex creep sustains a radially outward current of 
vortices and thereby a continuous internal torque, i.e. angular momentum transfer, from superfluid to crust.

A glitch occurs when the difference between the local superfluid velocities and the rotation rate of the crustal normal matter becomes too large to be sustained by pinning forces. Unpinned vortices will move in the radially outward direction from the rotation axis. This results in a sudden transfer of angular momentum to the crust normal matter, observed as the glitch. The sudden decrease $\delta\Omega$ in the superfluid rotation rate and increase in the crust rotation rate reduces the lag, temporarily stopping the creep process. There is thus a sudden decrease in the internal torque (which is mediated by the creep process) on the crust, while the pulsar torque continues to spin the crust down. This results in an increase of the observed spin-down rate of the crust. The glitch-associated decrease $\delta\Omega$ in the superfluid rotation rate is proportional to the total number of unpinned 
vortices and relates to the size of the observed spin-up $\Delta\nu$. The total angular momentum transferred from the superfluid to the crust in a glitch is proportional to both $\delta\Omega$ and the moment of inertia of the 
distinct superfluid regions, including creep regions and vortex-depleted regions, through which the unpinned vortices have moved.

The vortex creep model also has a linear regime, prevailing in some superfluid regions whose response explains the prompt exponential relaxation and holds information on the temperature and 
pinning parameters. In PSR J0537$-$6910, in addition to the nearly constant $\ddot{\nu}$ behaviour and the exponentials seen in 12 glitches, ten inter-glitch intervals show more complicated features. These are not 
addressed in the present paper, but can provide additional insights about the non-linear creep process and further estimates of the pinning parameters and local temperature. These  detailed applications of the vortex creep model to PSR J0537$-$6910 are deferred to a separate paper \citep{erbil25a}.

In addition to the recovering response of interior components to glitch-induced offsets, there can be permanent spin-down shifts. Persisting (but not necessarily permanent) shifts in the spin-down 
rate are seen for several pulsars. Using some of the Crab pulsar's largest glitches, permanent changes with an average magnitude $\langle\Delta{\dot{\nu}}_{\rm p}/\dot{\nu}\rangle=2.51(16) \times 10^{ - 4}$ \citep{lyne15}, have been indentified \footnote{Derived 
from \cite{lyne15}, their Eq. (6) and Table 3, where a truly persistent term must be delineated from an exponentially decaying term.}. It has already been noted by \citet{antonopoulou18} that the cumulative effect of persistent shifts associated with glitches could explain the low (under 3 in general) inferred long-term braking indices for several glitching pulsars \citep{espinoza17}, including the negative effective braking index of PSR J0537$-$6910 \citep{antonopoulou18}. In this work, we consider the existence of true permanent shifts, which remain once the part of the initial glitch step $\Delta\dot{\nu}$ in spin-down rate has fully recovered due to both linear and non-linear internal torques (which lead to exponential and power-law-like recoveries, respectively). Note these permanent shifts differ from the measured persisting changes $\Delta{\dot{\nu}}_{\rm p}$ often given in the literature, which typically refer to all components of $\Delta\dot{\nu}$ that do not recover exponentially. In most pulsars, measured $|\Delta{\dot{\nu}}_{\rm p}|$ will thus contain a part that recovers under the rather stable $\ddot{\nu}$ due to the non-linear internal torques, and will be larger than any true permanent shift. To make the distinction we shall use the notation $\Delta{\dot{\nu}}_{\rm per}$ to denote true permanent shifts, with $|\Delta{\dot{\nu}}_{\rm per}|\leq|\Delta{\dot{\nu}}_{\rm p}|$. 

The Crab pulsar's glitch behaviour, extensively discussed by \citet{lyne15} is unique. Post-glitch recoveries show exponential components but do not display the high $\ddot\nu$ 
inter-glitch behaviour observed in PSR J0537$-$6910, the Vela and other pulsars. Long-term growing exponentials (in $\dot{\nu}(t)$), with $\tau = 320$ days and 
initial amplitude $0.54 \times |\Delta\dot\nu_{\rm p}|$ > 0, settling to a persistent shift $\Delta\dot\nu_{\rm p} < 0$ were seen following the 1975, 1989 and 2011 glitches, as described by 
\citet{lyne15} (in their Eq.(6), and displayed in their Fig.3). They infer persistent shifts for 10 Crab glitches in their Table 3 (There is a mistake 
in Eq.(3), which should contain the term $|\Delta\dot\nu_{\rm p}|$ rather than $\Delta\dot\nu_{\rm p}$ to reflect the behavior shown in Fig.(3)). 
Persistent shifts were already noted in the 1975 glitch \citep{demianskiproszynski83} and in the 1989 glitch by \citet{lyne92} who also found the `wrong sign' long-term 
exponential. As the next glitch often arrives long after the longest exponential component with $\tau \cong 320$ days has settled to its 
asymptotic value, the return to steady-state seems to be achieved before the arrival of the next glitch. Thus $\Delta\dot\nu_{\rm p}$ in the Crab appear to be permanent. 
In terms of the vortex creep model, the non-linear response of the internal torque in the Crab pulsar has a more complicated signature than the constant $\ddot\nu$ response because of the high internal temperature \citep{{ACCP96}}. The observed behaviour led to the idea that permanent shifts occur simultaneously with Crab glitches, possibly associated with crust breaking and motion of crustal plates `
leading to formation of new vortex traps or vortex depleted regions; in this scenario the `wrong sign' exponential is explained qualitatively by plate motion towards the rotation axis, inducing inward vortex motion \citep{ACCP94, ACCP96,erbil19}. 

 Since PSR J0537$-$6910 is a young pulsar with similarly high spin and spin-down rate as the Crab, the temperature profile 
inside these neutron stars and the crustal and superfluid pinning stresses could also be similar for the two pulsars. It is, therefore,  plausible that PSR J0537$-$6910 glitches can also result in permanent shifts like those observed from the Crab pulsar, likely to be due to sustained structural changes inside the neutron star that permanently affect the rotational dynamics of some superfluid 
components. This could be common if the vortex unpinning events take place in conjunction with 
crust breaking and formation of pinned vortex traps in crustal superfluid components, and such models have been employed for PSR J1119$-$6127 \citep{akbal15}, PSR J1048$-$5832 
\citep{liu25} and the Crab pulsar \citep{erbil19}. Both for the Crab and for the Vela pulsars, model-dependent estimates of the time to the next glitch improve 
significantly when permanent shifts of magnitude  $\Delta{\dot{\nu}}_{\rm per}/\dot{\nu} \gtrsim 10^{ - 4}$ are assumed \citep{akbal17,erbil20}.  Note that whilst the inferred $\Delta{\dot{\nu}}_{\rm per}/\dot{\nu}$ is an order of magnitude larger for the Vela pulsar than the corresponding fraction in the Crab pulsar, the permanent $\Delta{\dot{\nu}}_{\rm per}$ shifts are similar since the Crab slows down about 25 times faster than Vela.

\section{Model}\label{sec:model}

Glitches in PSR J0537$-$6910 have typical magnitudes 
$\Delta\nu/\nu=\mbox{a few}\times10^{-7}$ and $\Delta\dot{\nu}/\dot{\nu}=\mbox{a few}\times10^{-4}$ and happen approximately every $\sim100$ days. In order to obtain the contribution to the inter-glitch $\ddot{\nu}$ of the external braking torque from the 
pulsar's magnetosphere, denoted hereafter $\ddot{\nu}_{\rm 0}$, and the equivalent braking index, the glitch contributions (in the form of both recovering and permanent components) should be assessed and removed. 
Our approach involves two ingredients: 

(i) Most inter-glitch intervals display a roughly constant  $\ddot{\nu}_{\rm obs,j}$ (the subscript j indicates this is for  the inter-glitch interval after glitch j), which can be measured from observations. This trend can be decomposed in two components, corresponding to the external and internal torque contributions, as $\ddot{\nu}_{\rm obs,j}=\ddot{\nu}_{\rm 0}+\ddot{\nu}_{\rm ig, j}$, in which $\ddot{\nu}_{\rm ig, j}$ reflects the superfluid response.

(ii) The total spin-down rate change $\Delta\dot{\nu}_{\rm j}$ contains a part that recovers during the inter-glitch interval after glitch {\em j}, denoted $\Delta\dot{\nu}_{\rm ig,j}$, and a persisting part that has not recovered when the next glitch occurs and might be permanent: $\Delta\dot{\nu}_{\rm per,j} \equiv \Delta\dot{\nu}_{\rm j} - \Delta\dot{\nu}_{\rm ig,j}$. The cumulative effect of frequent shifts that persist at the time of the next glitch could cause the effective negative braking index of PSR J0537$-$6910, as noted by \citet{antonopoulou18} and \citet{ferdman18}. 

The glitches and persistent changes do not appear associated with changes in the electromagnetic signature of the pulsar, such as profile shape or intensity changes \citep{ferdman18}, therefore any permanent shifts likely reflect structural changes in the interior of the neutron star, namely in the solid crust. 
The natural physical model for such sudden and permanent 
structural changes is the starquake model \citep{baym71}. 
However, the overall glitch activity cannot be understood in terms of crust-quakes alone (see e.g. \citet{antonopoulou22}) and requires the involvement of the internal superfluid, see e.g. the explanation in terms of vortex unpinning in \citet{ACCP93,alpar95}. Crust breaking associated with permanent shifts in the spin-down rate at the time of a glitch, may be what triggers the sudden vortex unpinning which in turn achieves the angular momentum transfer to the outer crust observed as the glitch. 

\begin{figure}
	\includegraphics[width=\columnwidth]{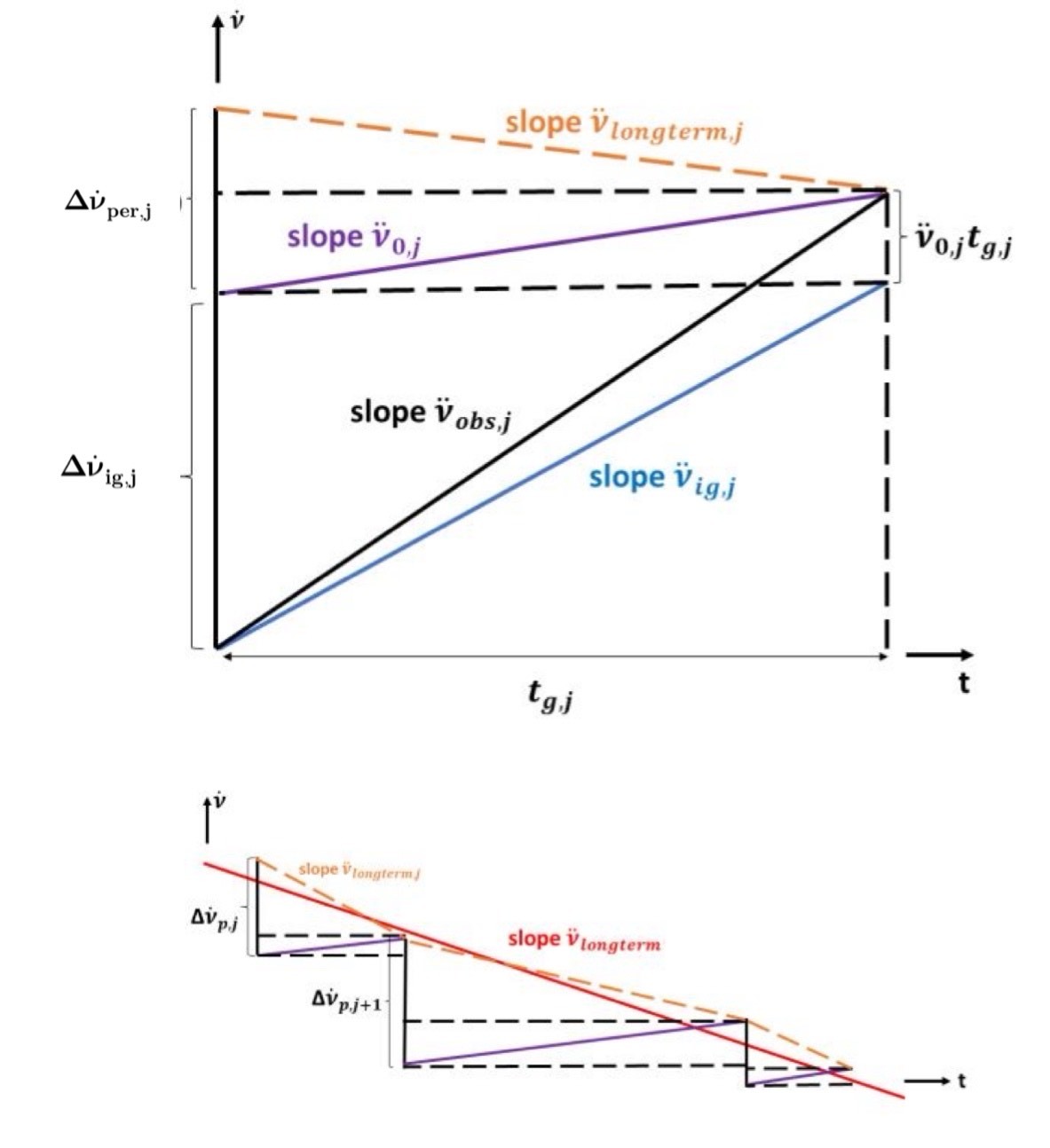}
    \caption {Upper panel: The model for inter-glitch evolution of $\dot{\nu}$ from glitch \textit{j} to glitch $j+1$. The black solid line depicts the observed inter-glitch 
    evolution.
    Assuming that 
    the recovery of the glitch offset in the internal torque is completed exactly at the time of arrival of the next glitch, the observed glitch step $\Delta\dot{\nu}_{\rm j}$ in spin-down rate is the sum of the permanent part $\Delta \dot{\nu}_{\rm per\ j} $ and the recovering part $\Delta \dot{\nu}_{\rm ig,j}$ under $\ddot\nu_{\rm ig,j}$ due to the internal 
    torque, as shown in blue. Subtracting the contribution of the internal torque from the total observed $\ddot\nu_{\rm obs,j}$ gives $\ddot\nu_{\rm 0,j}$, the slope of the 
    purple line, due to the external torque. The permanent shift $\Delta\dot\nu_{\rm per, j}$ is only partially counteracted by an amount $\ddot\nu_{\rm 0,j} t_{\rm g,j}$ 
    as shown by the purple line. The combined effect  results in a remaining decrease of $\dot{\nu}$, with an inferred slope $\ddot\nu_{\rm long-term,j}$ (orange line). 
    Lower panel: The long-term evolution over many glitches, shown in orange,  implies an effective negative $\ddot\nu_{\rm long-term}$ and negative effective braking index.}
\label{PSRJ0537braking}
\end{figure}

In this work, we make the assumption (discussed further in the following) that a glitch takes place when the relaxation from the previous glitch due to the internal torques is complete, i.e. the recovery under $\ddot{\nu}_{\rm ig, j}$ stops at the epoch of glitch $j+1$ and any remaining shifts in $\dot{\nu}$ are trully permanent. Making use of the $\theta$ (or Heaviside) function, $\theta(t) = 0$ for $t<0$, and  $\theta(t) =1$ for $t\geq 0$, we can then express the spin-down rate evolution over a time span including a total of $N_{\rm g}$ glitches, with recovering and permanent components, occurring at times $t_{\rm j}$:

\begin{align}
\dot\nu(t)&=\dot\nu_{0}+\ddot\nu_{0}t + \sum_{j=1}^{N_{\rm g}}\Delta\dot\nu_{\rm per,j}\theta(t-t_{j}) \nonumber\\ &+\sum_{j=1}^{N_{\rm g}}
\left(\Delta\dot\nu_{\rm  ig, j}+\ddot\nu_{\rm ig,j}(t-t_{\rm j}) 
\right)\;\theta(t-t_{\rm j})\theta(t_{\rm j+1}-t)\nonumber\\ &+\sum_{j=1}^{N_{\rm g}} 
+ [\Delta\dot\nu_{\rm  d,j} \exp (- (t-t_{\rm j})/\tau_{\rm  d,j})]\;\theta(t-t_{\rm j}),
\label{nudotev}
\end{align}
where $\dot\nu_{0}$ is a fiducial value and $\ddot\nu_{0}$ its (not directly observable) associated derivative based on the external dipole braking model. The first two terms then describe the evolution due to the external torque, whilst the rest are the glitch-associated changes. The third term, with $\Delta\dot\nu_{\rm per,j} < 0$, represents the permanent shifts, whilst the two last terms are the recovering components due to the internal superfluid torques: $\Delta\dot\nu_{\rm ig, j} <0 $ are the changes that recover with rate $\ddot\nu_{\rm ig,j}$ during the interval $ t_{\rm g,j} \equiv t_{\rm j+1}-t_{\rm j}$ until the next glitch $j+1$, and $\Delta\dot\nu_{\rm  d,j} < 0$ represent the exponentially recovering components. 

The total step increase in the 
spin-down rate observed at glitch $j$ is the sum of the persistent component and the recovering components:
\begin{equation}
\Delta\dot\nu_{\rm obs,j} \equiv \dot\nu(t_{\rm j+}) - \dot\nu(t_{\rm j}) = \Delta\dot\nu_{\rm per,j} + \Delta\dot\nu_{\rm ig,j} + [\Delta\dot\nu_{\rm  d,j}],
\label{nudottotal}
\end{equation}
where the epoch $t_{\rm j+}$ is immediately after the glitch at time $t_{\rm j}$.
The exponentially recovering components are not considered in this work, as they are not apparent in all glitches and, when measured, carry large uncertainties. Nonetheless, the inferred decay time-scales $\tau$ are under 40 d, so the later inter-glitch evolution will be dominated by the constant $\ddot\nu$ terms. The approach can be generalized to include exponentials in cases where it might prove necessary.

It is reasonable to expect that the time of completion of the inter-glitch recovery is close to the time of occurrence of the next glitch {\em 
on average}. This is because -- even though the actual glitch trigger 
can have a stochastic component -- physical conditions in the neutron star should be similar at the start of each glitch, given the regularity and similarity of their signatures in this pulsar. 
Moreover, it was observed for a sample of several pulsars that their average glitch waiting times scale with (and are in fact, very close to) their average $|\Delta\dot{\nu}|/\ddot{\nu}$, further supporting this idea \citep{lower21,liu25}, and \citet{ho20, ho22} also reported a similar correlation between $|\Delta\dot\nu(t_{\rm j})|/\ddot\nu_{\rm obs,j}$ and the time $t_{\rm g,j}$ from glitch {\em j} to the next glitch, $j+1$ for PSR~J0537$-$6910 specifically. 
We therefore make the aforementioned simplifying assumption that the inter-glitch recovery is completed exactly at the time when the next glitch arrives. 
The full recovery of $\Delta\dot \nu_{\rm ig,j}$ in the time interval $t_{\rm g,j} \equiv t_{\rm j+1} -  t_{\rm j}$ until the next 
glitch means that the contribution to $\ddot{\nu}$ of the internal torque is 
\begin{equation}
\ddot\nu_{\rm ig,j}  =  \frac{|\Delta\dot\nu_{\rm ig,j}|}{t_{\rm g,j}} 
\label{nuddotigj}\text{.}
\end{equation}

Unfortunately, the recovering and permanent parts $\Delta\dot \nu_{\rm ig,j}$ and $\Delta\dot\nu_{\rm per,j}$ of the glitch step in the spin-down rate cannot be separately inferred from the data, even under the assumption that $\Delta\dot \nu_{\rm ig,j}$ fully recovers exactly before the next glitch, because of the unknown external torque contribution. If, however, we assume a constant value of the permanent shift for all glitches, then we can estimate the relaxing part $\Delta\dot\nu_{\rm ig,j}$ of the glitch step $\Delta\dot\nu_{\rm obs,j}$ in glitch $j$ as
\begin{equation}
\Delta\dot\nu_{\rm ig,j}  =  \Delta\dot\nu_{\rm obs,j} - \langle\Delta\dot\nu_{\rm per}\rangle
\label{nudotcomp, nuddot0j}\end{equation}
and use it in Eq. (\ref{nuddotigj}) to approximate $\ddot\nu_{\rm ig,j}$. 
Subtracting this internally-driven $\ddot\nu_{\rm ig,j}$ from the total observed
$\ddot{\nu}_{\rm obs,j}$, gives   
$\ddot{\nu}_{\rm 0,j}$ due to the external pulsar torque: 
\begin{equation}
\ddot\nu_{\rm 0,j}  =  \ddot\nu_{\rm obs,j} - \ddot\nu_{\rm ig,j}
\label{nuddot0j}
\end{equation}
from which the braking index can be found. Equivalently, assuming a value for the braking index, the necessary permanent changes can be calculated.

The effects of the external and internal torques and glitch-induced step changes in $\dot\nu({\rm t}) $ are sketched in Figure \ref{PSRJ0537braking}. The total observed frequency second derivative $\ddot\nu_{\rm obs,j}$ is the slope of the black line whilst the blue line with slope 
$\ddot\nu_{\rm ig,j}$ and the purple line with slope $\ddot\nu_{\rm 0,j}$ represent the contributions of the internal torque and the external (pulsar) torque to the inter-glitch 
$\dot\nu$ evolution, respectively. The orange dashed line with slope $\ddot\nu_{\rm long-term,j}$ depicts the long-term evolution dominated by the effects of persisting shifts.
 
After many glitches, over the entire time span of the observations, equation (\ref{nudotev}) reduces to 
\begin{equation}
\dot\nu(t)=\dot\nu_{\rm 0}+\ddot\nu_{\rm 0}t+\sum_{j=1}^{N_{\rm g}}\Delta\dot\nu_{\rm per,j}\theta(t-t_{\rm j}).
\label{longterm} 
\end{equation}
The first sum in Eq. (\ref{nudotev}), which describes the recovering part of the glitch-induced steps in the spin-down rate, is zero based on our assumption of complete recoveries. 
Thus, it is the sum of the permanent shifts $\Delta\dot\nu_{\rm per,j} < 0$  over all glitches that results in the negative long-term slope of $\dot\nu(t)$, despite a positive 
$\ddot\nu_0$ arising from the external (pulsar) torque. 

\section{Results}
\label{sec:results}

In the following, we use a sample of 32 glitches for which $\ddot\nu_{\rm obs,j}$  and $\Delta\dot\nu_{\rm obs,j}$ are relatively accurately measured. We choose to exclude glitches with uncertainties larger than $50\%$ in either of these quantities. 
This criterion leaves 23 RXTE glitches and 9 NICER ones, whose properties are given in the first 3 columns of Table \ref{obsmod}. The average inter-glitch interval for this sample is
$\langle t_{\rm g}\rangle \cong $ 122 days, with an rms deviation of 12 days. 
For a nominal value of the underlying braking index, taken as $n=3$ in this instance, we use the observed parameters and Eqs. (\ref{nuddotigj}), (\ref{nudotcomp, nuddot0j}), and \ref{nuddot0j}) of the model, to calculate the values of the permanent shifts $\Delta\dot\nu_{\rm per}$ and  $\ddot\nu_{\rm ig}$ for these glitches, which are also presented in Table \ref{obsmod}.

If the long-term apparent braking index $n'$  is due 
to permanent shifts, we can calculate how big -- on  average -- they must be, to bring an actual braking index of $n$ due to the external torque down to the observed {\em negative} $n'$ value. The difference in $\ddot\nu$ yields -- from Eqs. (\ref{nudotcomp, nuddot0j}), (\ref{nuddotigj}), (\ref{nuddot0j}) and a given $n$ -- the average permanent shift needed: 
\begin{equation}
\langle\Delta\dot\nu_{\rm per}\rangle_{\rm n}  =  \langle(\ddot\nu_{\rm long-term} - \ddot\nu_{\rm n}) t_{\rm g}\rangle = (n'-n)\frac{{\dot\nu}^2}{\nu}\langle t_{\rm g}\rangle,
\label{avepern}
\end{equation} 
where $\ddot\nu_{\rm 0}$ is the expected value for a spin-down law with a constant braking index $n$. 

For the long-term $n'$ we use the value $ n' = - 1.234(9)$, corresponding to $\ddot\nu_{\rm long-term} = - 7.92(6) \times 10^{-22}$ Hz s$^{-2}$, measured by \citet{ho22} from a linear fit to the combined RXTE and NICER data. We, moreover, take a reference value $n = 3$, an average inter-glitch interval of $\langle t_{\rm g}\rangle \cong $ 122 days, and a typical $\dot{\nu}^2/\nu$, and find that the average permanent shifts must be about  $\langle\Delta\dot\nu_{\rm per}\rangle_{\rm n = 3} \simeq - 3 \times 10^{- 14}$ Hz s$^{-1}$. 
This result is not very sensitive to the chosen value of the braking index $n$; for a range of $1\leq n \leq 7$ the corresponding required average permanent shifts vary from $-1.5$ to $-5.5$ $\times 10^{-14}$ Hz s$^{-1}$. In general, the required average $\langle\Delta\dot\nu_{\rm per}\rangle$ for the theoretically predicted range of magnetosphere-driven braking models (see, e.g., \cite{melatos97,ruderman98,yan12,zhang22}) are typically a modest fraction of the total change $\Delta\dot{\nu}$ (quoted in Table \ref{obsmod}). 
An upper limit for the permanent shifts can be obtained from the persistent steps: after the first glitch, for which the exponential decay can be accounted for, the persisting change is found to be $\Delta\dot\nu_{\rm p, 0537} = -7.0\pm0.4\times 10^{-14}$ Hz s$^{-1}$  \citep{antonopoulou18}.
In the limiting case where the average permanent shifts are as large as the observed persisting one, the underlying braking index would be $9$.

For comparison with possible permanent shifts in other pulsars, the ratio $\langle\Delta\dot\nu_{\rm per}\rangle / \dot\nu$ may be the relevant quantity, as it gives an effective fractional moment of inertia of the component responsible. 
In the Crab pulsar, the average of the persistent steps reported by \citet{lyne15} for 10 glitches is $\langle\Delta\dot\nu_{\rm p}\rangle_{\rm Crab}\sim -9\times10^{- 14}$ Hz s$^{-1}$ 
%(-9.46 \pm 0.6) \times 10^{- 14}$ Hz s$^{-1}$, 
 which gives a fractional change of  $\langle\Delta\dot\nu_{\rm p}\rangle / \dot\nu|_{\rm Crab} = (3 \pm 2) \times 10^{-4} $. This value is rather close to the value we infer for PSR J0537$-$6910 under an assumed $n=3$:
$\langle\Delta\dot\nu_{\rm per}\rangle_{\rm n = 3}/\dot\nu|_{\rm 0537} = 1.44 \times 10^{ - 4}$. For the Vela pulsar, \citet{akbal17} calculated the required permanent shifts in a model-dependent way, by assuming the same glitch signature as in Eq. (\ref{nudotev}) and calculating the average $\langle\Delta\dot\nu_{\rm per}\rangle$ that minimizes the difference between observed and predicted (assuming full recovery) glitch epochs. 
They obtained $\langle\Delta\dot\nu_{\rm per}\rangle / \dot\nu|_{\rm Vela} = 2.32 \times 10^{-3} $, which is an order of magnitude larger than the Crab and PSR~J0537$-$6910 results\footnote{The difference could be because the older Vela pulsar has accumulated a larger network of vortex trap sites that trigger glitches \citep{cheng88,ACCP93}}. However, their inferred  $\langle\Delta\dot\nu_{\rm per}\rangle_{\rm Vela} = (-3.64 ) \times 10^{- 14}$ Hz s$^{-1}$ is very similar to the ones we find for  PSR~J0537$-$6910. Similarly to the upper bound we obtain on $|\Delta\dot{\nu}_{\rm per}|$ from observations of the first PSR~J0537$-$6910 glitch, a limit for the permanent changes in Vela can be found from the observed persisting shifts (once exponentials have been removed). Using the persisting changes measured by \citet{yu13} for 4 Vela glitches, we find the limit 
$\langle\Delta\dot\nu_{\rm per}\rangle / \dot\nu|_{\rm Vela} < \langle\Delta\dot\nu_{\rm p}\rangle / \dot\nu|_{\rm Vela} = (5 \pm 1) \times 10^{-3} $. It is thus conceivable that permanent changes as the ones we assume here occur in young glitching pulsars.  

These considerations provide proof of concept that the irregular rotational evolution of PSR J0537$-$6910 can be explained by a regular underlying braking index, for instance $n\sim1-7$, and the effects of its glitches alone. The high inter-glitch braking indices are a result of the internal superfluid torque that dominates until the next glitch, whilst the anomalous long-term negative $n'$ is the cumulative result of glitch-related, abrupt permanent shifts with physically plausible magnitudes that can be accommodated by the timing data. 

\begin{table*}
\label{fitpar}\centering\scriptsize
\caption{The glitch observables $\Delta\nu, \Delta\dot\nu, \ddot\nu_{\rm ig}, t_{\rm g}$ along with the values for persistent shift $\dot\nu_{\rm per}$ corresponding to a braking index 
$n=3$. The last column indicates the reference that glitch parameters are taken from. See the text for the criteria of chosen glitches and explanations of the model equations.}
   \begin{tabular}{cccccccc}
    Glitch Epoch (MJD). & $\Delta\nu$ ($\mu$Hz) & $ \Delta\dot\nu$ (10$^{-14}$\,Hz~\mbox{s$^{-1}$}) & $ \ddot\nu_{\rm ig}$ (10$^{-20}$\,Hz~\mbox{s$^{-2}$}) & $t_{\rm g}$ (days)  & $\Delta\dot\nu_{\rm per}\,(n=3)$ (10$^{-14}$\,Hz~\mbox{s$^{-1}$}) & Reference \\
    \hline
    
   51278 & 42.6(2) & -24.7(10.2) & 0.49(1) & 284(31) & -6.65 & \cite{antonopoulou18} \\
   51562 & 27.9(2) & -14.8(1.5) & 0.75(4) & 149(20) & -3.49 & \cite{antonopoulou18}\\
   51711 & 19.5(1) & -12.3(1.3) & 1.22(9) & 115(13) & -2.69 & \cite{antonopoulou18}\\
   51881 & 8.7(1) & -13.9(4.4) & 2.8(5) & 79(11) & -1.85 & \cite{antonopoulou18}\\
   52170 & 11.4(1) & -15.5(3.2) & 1.7(5) & 71(17) & -1.66 & \cite{antonopoulou18}\\
   52241 & 26.44(5) & -4.8(1.2) & 0.64(5) & 137(23) & -3.21 & \cite{antonopoulou18}\\   
   52545 & 26.1(1) & -9.2(4.0) & 0.72(2) & 186 (20) & -4.36 & \cite{antonopoulou18}\\
   52807 & 15.8(2) & -12.5(5.9) & 1.8(4) & 79(18) & -1.85 & \cite{antonopoulou18}\\
   52886 & 14.55(2) & -8.7(9) & 1.33(5) & 128(9) & -3.00 & \cite{antonopoulou18}\\
   53014 & 21.0(1) & -14.3(1.2) & 0.95(4) & 111.5(6.1) & -2.61 & \cite{antonopoulou18}\\
   53145 & 24.25(1) & -3.8(7) & 0.87(4) & 143(5) & -3.35 & \cite{antonopoulou18}\\
   53288 & 24.51(4) & -13.7(1.5) & 1.13(5) & 157(5) & -3.68 & \cite{antonopoulou18}\\
   53445 & 16.09(4) & -17.4(2.1) & 1.7(1) & 105(4) & -2.46 & \cite{antonopoulou18}\\
   53550 & 19.90(4) & -13.4(2.3) & 1.06(8) & 146(16) & -3.42 & \cite{antonopoulou18}\\
   53696 & 25.4(2) & -13.9(1.9) & 0.83(5) & 165(15) &-3.87 & \cite{antonopoulou18}\\
   53861 & 14.56(4) & -16.7(2.8) & 2.2(4) & 90.3(1.3) & -2.12 & \cite{antonopoulou18}\\
   54271 & 30.3(1) & -15.4(4.1) & 0.79(3) & 177(9) & -4.16 & \cite{antonopoulou18}\\
   54448 & 14.8(1) & -15.1(3.0) & 1.6(3) & 90(11) & -2.11 & \cite{antonopoulou18}\\
   54767 & 22.4(1) & -11.2(2.8) & 1.05(6) & 128(13) & -3.01 & \cite{antonopoulou18}\\
   54895 & 21.1(1) & -10.3(1.1) & 0.95(3) & 148(11) & -3.48 & \cite{antonopoulou18}\\
   55043 & 13.45(3) & -15.9(1.6) & 2.13(7) & 141(4) & -3.31 & \cite{antonopoulou18}\\
   55184 & 12.94(4) & -22.3(2.7) & 0.7(2) & 96(6) & -2.26 & \cite{antonopoulou18}\\
   55451 & 10.47(4) & -7.9(1.5) & 1.9(7) & 68(9) & -1.60 & \cite{antonopoulou18}\\
   58152 & 36.035(6) & -16.1(1) & 0.56(1) & 211(25) & -4.97 & \cite{ho20}\\
   58363 & 7.829(55) & -22.9(3.6) & 5.9(6) & 61(16) & -1.43 & \cite{ho20}\\
   58637 & 26.986(13) & -8.6(4) & 0.88(3) & 170(11) & -4.01 & \cite{ho20}\\
   58807 & 7.565(30) & -22.1(2.5) & 5.76(86) & 61(8) & -1.44 & \cite{ho20}\\  
   58868 & 24.038(84) & -24.4(4.7) & 1.06(8) & 125(8) & -2.95 & \cite{abbott21a}\\
   58993 & 0.426(9) & -0.77(35) & 1.0(1) & 56(6) & -1.32 & \cite{abbott21a}\\
   59049 & 8.457(22) & -13.1(1.4) & 3.6(8) & 54(8) & -1.27 & \cite{abbott21a}\\
   59351 & 12.27(3) & -7.9(2.2) & 1.6(1) & 103(10) & -2.43 & \cite{ho22}\\
   59454 & 16.60(1) & -17.1(4) & 1.0(4) & 68(16) & -1.60 & \cite{ho22}\\
 \hline
    \end{tabular}
\label{obsmod}
\end{table*}

\section{When is the Next Glitch Due?}\label{glitchpre}

Within our model, given a chosen $\left( \langle\Delta\dot\nu_{\rm per}\rangle\,,\,n\right)$ set of values obeying Eq. \ref{avepern}, one can deduct the contribution of the internal torque $\ddot{\nu}_{\rm ig}=\ddot{\nu}_{\rm obs} -  n {\,\dot{\nu}^2}/{\nu}$.  
If the last glitch of the pulsar has a measured $\Delta\dot\nu_{\rm obs,last}$ and we assume it involved a permanent shift of order $\langle\Delta\dot\nu_{\rm per}\rangle$, then the recovering part is estimated as $\Delta\dot{\nu}_{\rm ig,last}\simeq \Delta\dot{\nu}_{\rm obs,last} - \langle\Delta\dot\nu_{\rm per}\rangle_{\rm n}$. From these we can estimate the time it takes for a full recovery :   
 
\begin{equation}
t_{\rm g,next,n} \cong \frac{|\Delta\dot\nu(t_{\rm ig,last})|}{\ddot\nu_{\rm ig,last}} \cong \frac{\Delta\dot{\nu}_{\rm obs,last} - (n' - n)(\dot\nu^2/\nu)\langle t_{\rm g}\rangle}{\ddot{\nu}_{\rm obs,last} -  n \dot\nu^2/\nu},
\label{tnext}
\end{equation}
in which we substituted $\langle\Delta\dot\nu_{\rm per}\rangle_{\rm n}$ from Eq.(\ref{avepern}).
This time-scale approximates the waiting time until the next glitch. 

Another estimate of the time to the next glitch can be found using the vortex creep model.  
The sudden motion of unpinned vortices in the glitch transfers angular momentum from the superfluid to the crust. Equating the angular momentum transfer from the superfluid to the angular momentum change of the crust (manifested as the observed glitch), and relating the moments of inertia of the vortex creep and vortex-depleted superfluid regions  with the step in the spin-down rate gives \citep{alpar84}
\begin{equation}
\Delta\nu_{\rm j} = \frac{\Delta\dot\nu_{\rm ig, j}}{\dot\nu}(1/2 + \beta_{\rm j})\delta\nu_{\rm j} \cong \frac{\Delta\dot\nu_{\rm ig,j}}{\dot\nu}(1/2 + \beta_{\rm j})|\dot\nu|t_{\rm g,j},
\label{angularmomentum}
\end{equation}
where 
$\delta\nu_{\rm j}$ is the decrease in the superfluid rotation frequency resulting from the sudden vortex motion driving the spin-up $\Delta\nu_{\rm j}$. In the vortex creep model $\Delta\dot\nu_{\rm ig, j}/\dot\nu$ is 
interpreted as the fractional moment of inertia, denoted $I_A/I$, of the vortex creep regions where vortices 'creep' against background of pinning sites, and $\beta_{\rm j}$ is a model 
parameter representing the ratio of the moments of inertia of the vortex-depleted superfluid regions (whose fractional moment of inertia is denoted by $I_B/I$ in the vortex creep 
literature) to those regions that contain creeping vortices \citep{cheng88,alpar06}.

The next glitch is 
expected to occur {\em roughly} when the offset lag between the superfluid and the crust-normal matter rotation rates is recovered by the spin-down of the crust-normal matter. This 
leads to the estimation $t_{\rm g,j}\cong \delta\nu_{\rm j} /|\dot\nu|$ and to the second, approximate, equality in Eq. (\ref{angularmomentum}). Using Eqs. (\ref{nuddotigj}), 
(\ref{nuddot0j}) and (\ref{avepern}) and assuming that the parameter $\beta_{\rm j} = \beta$, a constant representative value for all glitches, we obtain
\begin{equation}
\frac{\Delta\nu_{\rm j}}{\ddot\nu_{\rm ig, j} - n\dot\nu^2/\nu} \cong (1/2 + \beta){t_{\rm g, j}}^2.
\label{creepmodelestimate}
\end{equation}
In the case of PSR J0537$-$6910, assuming $n=3$ and the glitches in Table \ref{obsmod}, a fit of Eq. (\ref{creepmodelestimate}) returns $\beta=18.4$. This is much larger than inferred 
$\beta$ parameters for other pulsars, which was found to vary from as low as $\lesssim0.5$ for the Crab and two more pulsars \citep{erbil19,erbil21} up to   $\cong5$ for other sources  \citep{liu25,liu24}. The difference in inferred $\beta$ for this pulsar poses some question for the interpretation within the creep model, but in principle $\beta$ could be used in Eq. (\ref{creepmodelestimate}) together with $\Delta\nu_{\rm last}$ and $\ddot\nu_{\rm ig, last}$ following the last glitch, to approximate $t_{\rm g, next}$. 

Even though none of the current observed \citep{lower21, liuyang24,ho20} or theoretical correlations provides a very tight estimate of the time of a future glitch, the predictability of glitches in this pulsar can be used for targeted observations. The strong correlation between $\Delta\nu_{\rm obs,j}$ and $t_{\rm g,j}$ that is well established for this pulsar \citep{middleditch06,antonopoulou18} gives a typical uncertainty window around 10-20 d for the time of the next glitch, assuming good coverage to accurately measure $\Delta\nu$ of the previous glitch \citet{ho20, ho22}. 

\section{Discussion and Conclusions}\label{sec:conc}

We have studied the timing behaviour of the unique source PSR J0537$-$6910 in terms of its many glitches, using published timing fits  
\citep{antonopoulou18, ho20,ho22, abbott21a} which measure a high, relatively constant, inter-glitch 
frequency second derivative $\ddot{\nu}>0$ and an effective $\ddot{\nu}_{\rm long-term}<0$ inferred from the decrease of $\dot{\nu}$ in the long-term. We propose that the apparent long-term negative $\ddot{\nu}_{\rm long-term}$ of PSR J0537$-$6910 arises from permanent, non-relaxing steps in the spin-down rate associated with each glitch, 
like the persistent shifts observed in the Crab pulsar \citep{lyne15} and inferred for the Vela pulsar \citep{akbal17} and for PSRs J1119-6127 \citep{akbal15} and J1048--5832 
\citep{liu25}. The separate effects of the external torque, the internal torque, and that of permanent $\dot{\nu}$ changes cannot be extracted from the timing evolution. Making a justified approximation, however, that a glitch occurs when the recovery (both exponential and power-law-like) from the previous glitch is complete, we find a relationship between the braking index $n$ of the external torque and the averaged required permanent changes that will produce the effective $\ddot{\nu}_{\rm long-term}<0$ over the years. 
Adopting an underlying pulsar torque of braking index $n\sim 3$ yields the permanent shift values $\Delta\dot{\nu}_{\rm per} \sim 10^{-14}$ Hz s$^{-1}$  listed in Table \ref{obsmod}, with an average value $\langle\Delta\dot\nu_{\rm per}\rangle_{\rm n = 3}= - 2.86 \times 10^{- 14}$ Hz s$^{-1}$, of similar order of magnitude as those inferred for the Crab and Vela pulsars. The fractional permanent changes inferred for the Vela pulsar are an order 
of magnitude larger, interpreted as indicating a larger connected network of vortex traps in the Vela pulsar \citep{akbal17}. 

Thus, we have shown that internal torques combined with permanent shifts 
and external pulsar torques with braking indices not far from the canonical value of $n = 3$, suffice to explain the timing behavior of PSR J0537$-$6910. 

We have based our discussion of the time between glitches on the expectation that the next glitch occurs at about the time when internal torques have recovered to pre-glitch equilibrium conditions. However, the requirement of permanent shifts in the spin-down rate suggests possible re-arrangement of the crust 
geometry in relation to the glitches. Crust breaking could be initiating local vortex unpinning, which then evolves into a large vortex avalanche and observed glitch, when the internal conditions are appropriate. \citet{middleditch06} discuss possible crustquakes in PSR~J0537$-$6910 in connection to its glitches, and crust failure is often suggested as a possible glitch trigger to explain the scale-invariant glitch size distributions seen in some pulsars like the Crab. Conversely, additional stresses to the crust by the pinned vortices could initiate both a crustquake \citep{ruderman91} and (subsequent) large-scale unpinning when the superfluid lag grows large. Breaking and moving of a plate has also been invoked to explain some glitch signatures (and associated emission changes) in the highly magnetised pulsar PSR~J1119$-$6127 \citep{{akbal15, akbal18}}. Such local breaking requires a non-axisymmetric treatment of elastic deformations along the lines 
of recent work by \citet{GangwarJones24} on mountains on neutron stars. A detailed 
discussion of the stresses induced by vortex pinning, the building of a vortex avalanche starting with vortices unpinned from a specific broken crustal plate, 
the motion of unpinned vortices under dissipative interactions with electrons and their encounters with vortex traps (clusters) to unpin more vortices, will be the subject of a separate 
paper \citep{erbil25b}. Here we just note in passing that the size of a broken plate, like the maximum height of a mountain above the neutron star surface, determines a local, non-axisymmetric deformation, which can  
source gravitational waves emission \citep{chamel08,GangwarJones24}.

The large inter-glitch braking indices seen for PSR~J0537$-$6910 \citep{antonopoulou18,ferdman18} have also evoked the idea of possible emission of gravitational waves arising from internal fluid 
motions from r-modes \citep{andersson18} or crustal deformations \citep{cutler02}, which give predicted braking indices $n = 7$ and $n = 5$, respectively. Such contributions to the spin-down evolution are not ruled out by the current analysis. The expected 
gravitational wave strain amplitude from r-modes was estimated as $h \sim 2-3 \times 10^{- 26}$ \citep{andersson18}. Initial searches with the third LIGO/Virgo survey have failed to detect gravitational 
radiation associated with PSR J0537$-$6910 \citep{fesik20, abbott21a, abbott21b}. The most recent fourth LIGO-Virgo-KAGRA observing run \citep{abac25} that covered the timespan containing recent 
glitches of PSR J0537$-$6910 reported by \citet{ho20}, placed a constraining upper limit $h \lesssim 10^{ - 26}$ for the gravitational wave strain amplitude. 
It remains imperative to discover if gravitational wave emission contributes to the braking of PSR J0537$-$6910, which -- together with continuous monitoring in the X-rays -- can significantly alter the inferences for the internal and magnetospheric torques of this pulsar.
  
Frequent monitoring of PSR~J0537$-$6910 is also vital to better characterise its glitches for use as input on theoretical calculations. Model-dependent analyses and simulations can help resolve the ambiguity in separating any permanent shifts in spin-down rate and the ongoing power-law recovery due to internal torques. In the framework of the vortex creep model, this can lead to estimates of the time to the next glitch; a complete analysis of the superfluid recovery 
contributing to the constant $\ddot\nu$ behavior, as well as the prompt exponential post-glitch relaxations can yield information on the 
moments of inertia of the interior components involved, on the temperature, and on superfluid vortex pinning. Likewise, the reported correlations between glitch size and the time interval 
to the next glitch \citep{antonopoulou18,ferdman18,ho20,liuyang24} is used to  estimate superfluid moments of inertia 
and the angular momentum exchange at glitches \citep{antonopoulou18, erbil25a}. 

As a final remark we would like to call attention to possible glitch induced variations in pulse profile or emission properties. A handful of pulsars exhibited 
pulse profile changes associated wtih glitches [see, for example, Table 1 in \citet{zhou23}]. In particular, the best glitch resolved to date, the 2016 Vela glitch, 
displayed clear polarization level variations and a nulling pulse around the spin-up \citep{palfreyman18}. These ephemeral changes might also point to a glitch-accompanying crustquake, which can move the footpoints of the magnetic field lines close to the polar cap \citep{akbal15} and release seismic energy into the 
magnetosphere \citep{bransgrove20}. While these short-lived and ephemeral changes observed at radio wavelengths are expected to arise from perturbations in the emission regions just above 
the polar cap, the origin of the X-rays likely lies in the outer gaps at much higher altitudes in the magnetosphere \citep{philippov22}. Because PSR J0537$-$6910 is only seen in X-rays, detecting a glitch associated pulse profile variation could be more difficult, however the predictability of a forecoming glitch in PSR~J0537$-$6910 can be used to plan future targeted observations.

\begin{acknowledgements} 
 
We acknowledge the Scientific and Technological Research Council of Turkey (T\"{U}B\.{I}TAK) for support under the grant 117F330. EG is supported by National Natural Science Foundation of China (NSFC) 
programme 11988101 under the Foreign Talents Grant QN2023061004L. D.A. acknowledges support from an EPSRC/STFC fellowship ( EP/T017325/1). 
 We thank the anonymous referee for many constructive comments and suggestions, that helped to improve the clarity of this article.
 
\end{acknowledgements}

\end{document}